\documentclass[12pt]{iopart}
\usepackage{amsthm,amssymb,url,enumitem,bm}
\expandafter\let\csname equation*\endcsname\relax
\expandafter\let\csname endequation*\endcsname\relax
\usepackage{amsmath}
\usepackage[noabbrev,capitalize]{cleveref}
\crefname{equation}{}{}
\usepackage{graphicx}

\usepackage{enumitem}
\usepackage{algorithm}
\usepackage{algpseudocode}
\usepackage{scalerel}
\usepackage{xcolor,soul}

\usepackage{subcaption}
\usepackage{eucal}
\usepackage{xspace}
\newcommand{\deleted}[1]{}
\newcommand{\later}[1]{}
\newcommand{\almset}{ALM\_\,set\xspace}
\newcommand{\almsetpl}{ALM\_\,set\textsuperscript{+}\xspace}
\renewcommand{\ss}{\bm{s}}
\renewcommand{\ll}{\bm{\lambda}}

\renewcommand{\algorithmicrequire}{\textbf{\small Input:}}
\renewcommand{\algorithmicensure}{\textbf{\small Output:}}

\begin{document}
\title[Logical qubit implementation: augmented Lagrangian approach]{Logical qubit implementation for quantum annealing: augmented Lagrangian approach}
\vspace{10pt}	

\author{Hristo N.\ Djidjev}

\address{
	Institute of Information and Communication Technologies, Bulgarian Academy of Sciences, Sofia, Bulgaria;\\
	Los Alamos National Laboratory, CCS-3, Los Alamos, NM 87545, USA
 }
\vspace{10pt}
	
\begin{abstract}
	Solving optimization problems on quantum annealers usually requires each variable of the problem to be represented by a connected set of qubits called a logical qubit or a chain. Chain weights, in the form of ferromagnetic coupling between the chain qubits, are applied so that the physical qubits in a chain favor taking the same value in low energy samples. Assigning a good chain-strength value is crucial for the ability of quantum annealing to solve hard problems, but there are no general methods for computing such a value and, even if an optimal value is found, it may still not be suitable by being too large for accurate annealing results. In this paper, we propose an optimization-based approach for producing suitable logical qubits representations that results in smaller chain weights and show that the resulting optimization problem can be successfully solved using the augmented Lagrangian method. Experiments on the D-Wave Advantage system  and the maximum clique problem on random graphs show that our approach outperforms both the default D-Wave method for chain-strength assignment as well as the quadratic penalty method.
\end{abstract}

\vspace{2pc}

\section{Introduction}
Quantum annealing (QA) is a method for solving optimization problems on special quantum devices that is targeting NP-hard problems that are difficult to solve on conventional computers \cite{kadowaki1998quantum}. The largest commercially available quantum annealers, which are designed by the company D-Wave Systems, have thousands of qubits, and can propose high-quality solutions of quadratic optimization problems of the type
\begin{equation}
	\mbox{minimize} \; \mathit{Is}(\ss) = \sum_{i<j}J_{ij}s_is_j+\sum_ih_is_i,\label{eq:Ising1}
\end{equation}
called \textit{Ising problem}, where $\ss=(s_1,\dots,s_N)$ and $s_i\in\{-1,1\}$ are the variables. When variables of  problem \cref{eq:Ising1} are are restricted to $s_i\in\{0,1\}$, the corresponding formulation is called a \textit{quadratic unconstrained binary optimization} (\textit{QUBO}) problem. The two formulations can easily be converted into each other by using a linear transformation of the variables, but D-Wave hardware natively implements the Ising version.  Problem \cref{eq:Ising1} is NP-hard to solve and many other NP-hard problems can be easily reformulated as Ising and QUBO problems \cite{Lucas2014}.

In order to solve an Ising problem \cref{eq:Ising1} on a D-Wave quantum annealer, the coefficients of the problem are sent to the device, which solves it multiple times using QA and return the obtained solutions called \textit{samples}, which are analyzed in postprocessing, and the sample resulting in a lowest value of the Ising function is chosen as a solution. This is justified by the fact that the QA itself takes a small portion of the total solution time, which also includes the time to map the problem on the hardware and to program it. 

\begin{figure}
	\centering
	\includegraphics[width=0.45\textwidth]{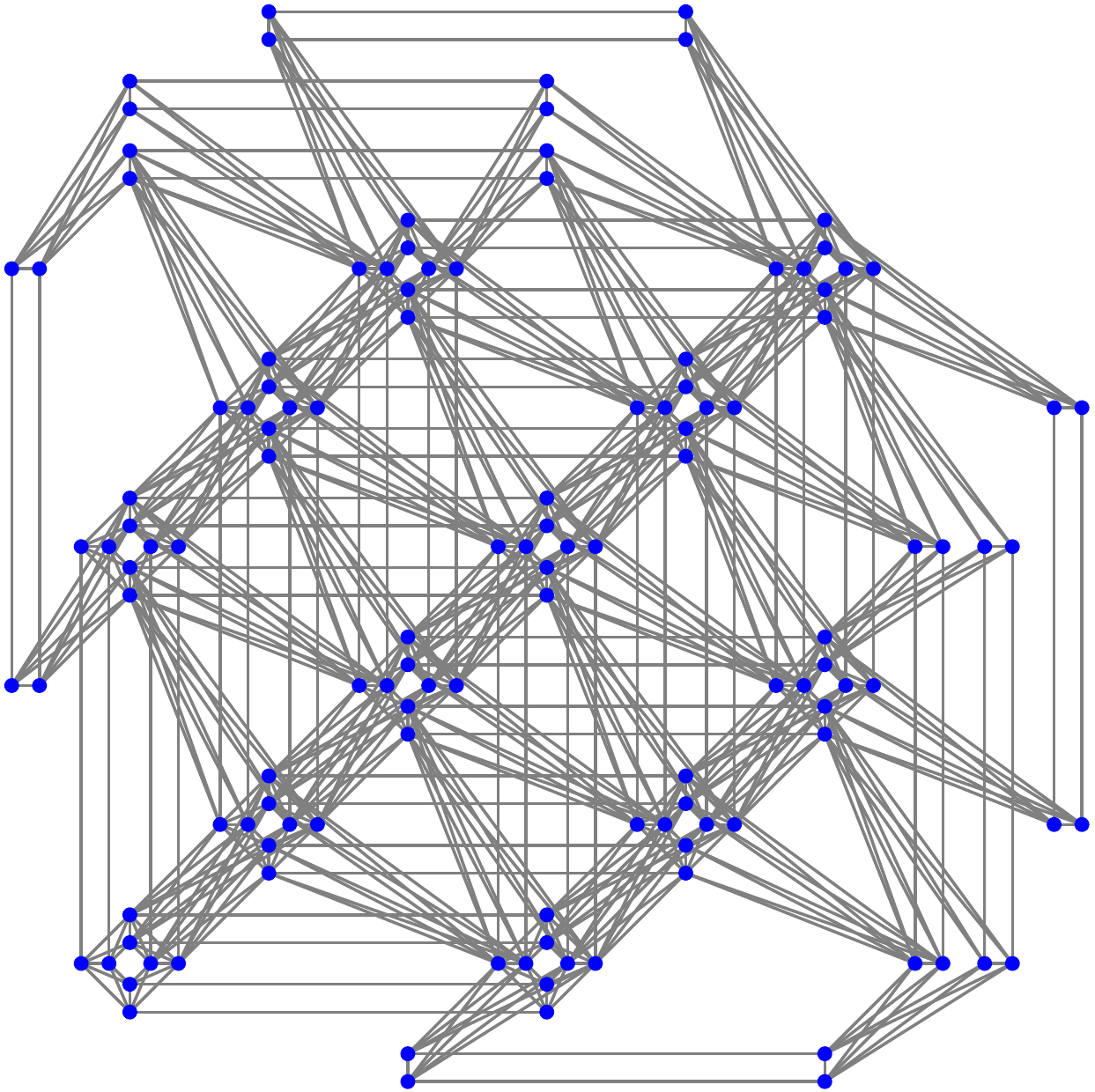}
	\caption{The connection pattern of a Pegasus graph, $P(3)$, of order 3. Blue dots are qubits and connecting edges are couplers. \texttt{Advantage\_system4.1} computer, used for experiments in this paper, is based on the $P(16)$ Pegasus graph and has 5627 working qubits and 40279 working couplers (some qubits and couplers are disabled due to defects).}\label{fig:pegasus_3}
\end{figure}

For mapping the Ising problem on the quantum processing unit (QPU), the coefficients $h_i$, called \textit{linear biases}, have to be mapped to distinct qubits of the QPU, while each coefficient $J_{ij}$, called a \textit{quadratic bias}, has to be mapped to a \textit{coupler} linking the qubits corresponding to $i$ and $j$. Such a coupler may not exist in the QPU, however, as the qubits in current D-Wave devices are not fully connected. \cref{fig:pegasus_3} shows the connection pattern of the Pegasus graph of the current largest quantum annealer, the Advantage\textsuperscript{\texttrademark} system, for which the average number of couplers for a qubit is about 15. In contrast, for many optimization problems, the graph corresponding to the quadratic biases is very dense, meaning that a nonzero coupler $J_{ij}$ exists for most of the possible pairs $(i,j)$. To address this issue, the Ising problem is minor-embedded into the QPU. Specifically, consider the graph with vertices $\{1,\dots,N\}$ and undirected edges $\{(i,j)~|~J_{ij}\neq 0\}$, which we refer to as the \textit{Ising graph}. Define as a \textit{hardware graph} the graph with vertices the qubits of the QPU and an edge between each pair of vertices whose corresponding pair of qubits are connected by a coupler. Then, a \textit{minor-embedding} of the Ising graph into the hardware graph is a mapping $M$ of the vertices of the Ising graph into connected subsets of vertices of the hardware graphs that are mutually disjoint and, for each edge $(i,j)$ of the Ising graph, there is an edge $(i',j')$ of the hardware graph such as $i'\in {M}(i)$ and $j'\in {M}(j)$. The sets of qubits corresponding to the sets vertices ${M}(i)$ are called \textit{logical qubits} or \textit{chains} for short. 
Finding such an embedding is NP-hard in general, but good heuristic algorithms are available \cite{choi2008minor,choi2011minor}.

In order for the set of qubits corresponding to a logical qubit to faithfully represent a single qubit, the qubits should be required to take the same value at the end of the QA process. 
For that end, each coupler $(i,j)$ from a chain is assigned a negative bias, say $-\mu$, favoring physical qubits $i$ and $j$ to take the same value in a minimization problem. The magnitude $\mu$ of that bias, controlled by D-Wave's parameter \texttt{chain\_strength}, is very important. If $\mu$ is too low, then it will not be able to enforce that the physical qubits in a chain take the same value. If $\mu$ is too large, then the accuracy of the representation of the Ising problem coefficients in the hardware will be reduced since, after normalization to the ranges accepted by the hardware, some problem coefficients may become too small for the limited resolution of the QPU. Picking a suitable value for the chain strength is usually a trial-and-error process and, although the D-Wave software offers a heuristic based method, called \texttt{uniform\_torque\_compensation}, for setting it, it doesn't always work well, resulting in so-called \textit{broken chains}, or logical qubits whose physical qubits have obtained both values $+1$ and $-1$. Such broken chains are resolved in postprocessing by a heuristic method such as the D-Wave's default method \texttt{majority\_vote}, which we use in our experimental analysis.
For more advanced methods for resolving broken chains, see \cite{pelofske2020advanced}.

Getting broken chains is a major obstacle to improving the accuracy and scalability of the quantum annealers \cite{coffrin2019challenges}. In this paper, we propose an optimization-based approach for finding good representations of the chains that uses not only quadratic biases on the couplers, as the previous methods, but also linear ones on the qubits. For this end, we represent the problem of minimizing an Ising function \cref{eq:Ising1} under the condition that all qubits in a chain should take the same value as a \textit{constrained optimization problem}. However, the Ising model \cref{eq:Ising1} does not allow constraints, so the constraints have to be added to the objective function. The typical method for solving constrained problems on D-Wave is the penalty method, which adds a constraint $\bm{a}\ss=\bm{b}$ as a quadratic penalty $\mu/2(\bm{a}\ss-\bm{b})^2$. In the case of logical qubit representation, this results in using negative bias $-\mu$ on the chain's couplers, and hence the penalty method produces a similar representation of the logical qubits as the one discussed above and suffers similar drawbacks. We show in this paper that better results can be achieved if the augmented Lagrangian method is used instead of the penalty one. The augmented Lagrangian method, which we discuss in more detail in a next section, adds a linear Lagrangian term $\ll^T(\bm{a}\ss-\bm{b})$ to the quadratic penalty, and uses an iterative procedure to find optimal or near-optimal values for $\ll$ and $\mu$. 

The main contributions of this paper are the following.
\begin{enumerate}[label=(\roman*)]
	\item We show that a good representation of the logical qubits can be constructed by defining both linear and quadratic biases and using the augmented Lagrangian method to find the values of such biases.
	\item We test our implementation on the maximum clique problem, an important NP-hard problem, and on graphs of up to 100 vertices.
	\item We show that our implementation outperforms both the default D-Wave representation of logical qubits as well as the quadratic penalty method.
	\item We show that one can optimize and store values of the augmented Lagrangian coefficients for a whole class of graphs and then use them to solve problems on any graph of the class.
\end{enumerate}

The rest of the paper is organized as follows. In \cref{sec:previous}, we overview previous results related to this work. In \cref{sec:methods}, we describe our methods and present the  pseudocodes of their implementation. In \cref{sec:results}, we analyze the proposed implementation of logical qubits and compare it with other algorithms. Finally, we conclude with a summary and a list of open problems.

\vspace{1cm}
\section{Previous work}\label{sec:previous}
Several studies have focused on empirically determining good chain strength values for certain problems.
King and McGeoch \cite{king2014algorithm} consider the algorithm-engineering aspects of solving optimization problems on quantum annealers including searching for a best value for the chain-strength parameter, which they empirically analyze on a subclass of 3-SAT instances. 
Pudenz \cite{Pudenz2016} uses empirical analysis to compare different strategies for setting parameters of logical qubits such as the distribution of the Ising problem biases on logical qubits and couplers and setting chain weights.
Grant and Humble \cite{grant2022benchmarking} benchmark  on a set of instances of portfolio optimization problems the probability of chain-breaks and how they are impacted by the specific embedding.
Lee \cite{Lee2022} use simulated annealing and exact optimization methods to estimate optimal chain strength  for hypothetical 2D $L\times L$ architectures for $L=4$, $6$, and $8$, where a third dimension is added to the 2D architecture to allow chains between the main qubits.
Venturelli et al. \cite{venturelli2015quantum} study optimal parameter settings on fully-connected Ising models corresponding to the Sherrington–Kirkpatrick spin-glass model. They establish that, for their problem, the optimal chain strength scales as $\sqrt{N}$, where $N$ is the number of the variables. For general Ising problems, the same method implies that the optimal chain strength scales proportionally to $\sigma\sqrt{N}$, where $\sigma^2$ is the variance of the quadratic couplers, which formula is used in the D-Wave software. Such a setting for the chain strength, along with the anneal duration and the minor embedding parameters, were analyzed in \cite{Raymond2020} for spin glasses and channel communication problems, and interdependence and strategies for tuning these three parameters were proposed. 

\section{Methods}\label{sec:methods}
We start this section by formulating  the logical qubits representation problem as an optimization problem and describe the generic quadratic penalty and augmented Lagrangian methods.

\subsection{Representing logical qubits as constraints}
	An alternative way of thinking about enforcing the condition that all physical qubits representing a given logical qubit $Q$ take the same value when solving an optimization problem $P$ is to consider it as an (additional) constraint added to $P$. 	In the case $P$ is a constrained problem, one can first convert it into an unconstrained one using the method described in the previous section.
	Hence, we can assume without loss of generality that $P$ is itself an Ising problem of the type \cref{eq:Ising1}, i.e., unconstrained.
	Assume a minor embedding for $P$ maps each variable $s_i$ onto a set $C_i=\{ s_i^1,\dots s_i^{k_i}\}$, where $s_i^k$ are the physical qubits, and let the set of active couplers joining pairs of vertices from $C_i$, which we call \textit{chain couplers},  be $E_i$. We assume that the chains, i.e., each graph $(C_i,E_i)$, is connected.
	Define ${\cal E}=\bigcup_i E_i$
	Denote by $C_{ij}$ the set $\{(k,l)\in {\cal E}~|~s_i^k\in C_i,\; s_j^l\in C_j\}$, i.e., the set of couplers between the chains $C_i$ and $C_j$, which we call a  \textit{logical coupler}.
	Replacing each $s_i$ by $C_i$, $i=1,\dots,N,$ in $\mathit{Is}(\ss)$ from \eqref{eq:Ising1}, we get the Ising function
	\begin{equation}
		\mathit{Is}'(\bm{s'}) = \sum_{i<j}\sum_{(k,l)\in C_{ij}}J_{ij}^{kl}s_i^ks_j^l+\sum_{i}\sum_{s_i^k\in C_i}h_i^ks_i^k,\label{eq:Ising2}
	\end{equation}
	where $J_{ij}^{kl}$ and $h_i^k$ are appropriately chosen coefficients and $\bm{s'}$ is the vector of all variables $s_i^k$, $k\in C_i$, $1\leq i\leq N$. The \textit{uniform spreading approach} \cite{Raymond2020}, used by D-Wave's Ising embedding tool, which we also use in the paper, assigns equal weights for each logical qubit and coupler. Hence, we define
	\begin{equation}
		J_{ij}^{kl} = J_{ij}/|C_{ij}|,\quad h_i^k=h_i/|C_i|.\label{eq:equal_weights}
	\end{equation}
	
	For enforcing that all variables $s_i^k$ in a logical qubit take the same value in an optimal solution, we can add the constraints
	\begin{equation}
		s_x=s_y \mbox{ for each $(s_x,s_y)\in \bigcup_iE_i$,}\label{eq:constraint}
	\end{equation}
	i.e., that the qubits at the endpoints of each chain coupler should take the same value. Since the graph $(C_i,E_i)$ is connected, \eqref{eq:constraint} ensures that all variables corresponding to any given chain will take the same value.
	
	However, the resulting problem 
		``\mbox{minimize $\mathit{Is}'(\bm{s'})$ subject to \eqref{eq:constraint}}'' 
	cannot be solved on D-Wave, as it is, since it is not unconstrained. To convert it into an unconstrained one, \eqref{eq:constraint} is usually added to the objective function as a penalty term, resulting in a problem
	\begin{equation}
		\mbox{minimize $\mathit{Is}'(\bm{s'}) + \varPi(\bm{s'})$},\; \bm{s'}\in\{-1,1\}^n,\label{eq:unconstrained1}
	\end{equation}
	where $\varPi(\bm{s'})$ is a \textit{penalty function} that takes a value zero, if all constraints are satisfied, and a positive value large enough to prevent  assignments that violate some of the constraints to be an optimal solution, otherwise. In the next subsection, we will discuss two methods for defining a suitable function $\varPi$.

\subsection{Penalty and augmented Lagrangian methods for constrained optimization}
The most common constraints for binary optimization problems are  equalities and inequalities between variables. We consider here domain ${\cal D}=\{-1,1\}^N$ and equality-constrained problems since this is the type of our problem \cref{eq:unconstrained1}, but the described methods was originally designed for continuous domains and it can be generalized for inequality-constrained or mixed problems. We consider in this subsection the constrained problem
\begin{equation}
	\mbox{minimize $f(\ss)$ \ subject to $c(\ss)=0,\; \ss\in {\cal D}$.}\label{eq:problem}
\end{equation}
\subsubsection{Penalty method}
The quadratic penalty method (PM) uses a penalty function $\varPi(\ss)=\mu/2\,||c(\ss)||^2$, transforming \cref{eq:problem} into the unconstrained problem
\begin{equation}
	\mbox{minimize $f(\ss)+\dfrac{\mu}{2}\,||c(\ss)||^2,\; \ss\in {\cal D}$}.\label{eq:problem2}
\end{equation}
Choosing the \textit{penalty factor} $\mu>0$ large enough will enforce $c(\ss)=0$ in each optimal solution of \eqref{eq:problem2} (assuming \cref{eq:problem} does have a feasible solution) thereby producing an optimal solution of \eqref{eq:problem}. But if the penalty factor is chosen too large, that will result in some very small in absolute value coefficients of the Ising problem submitted to the quantum annealer, since all coefficients are normalized in order to satisfy hardware-imposed limits. Combined with the analog nature of the quantum annealer and the finite digital-to-analogue converter quantization step size, such large values of $\mu$ would negatively affect the quality of the solution of problem \eqref{eq:problem2}. Even in the continuous case, large values of $\mu$ result in ill-conditioning. 
Therefore, finding a suitable penalty factor is important for the success of the method. 

We will use in our implementation of the penalty method an iterative procedure that computes a suitable penalty factor, which starts with a small value of $\mu$ and increases that value in each iteration until all constraints are satisfied. A pseudocode for the method is given in Algorithm~\ref{alg:penalty}.

\begin{algorithm}
	\caption{Quadratic penalty method}\label{alg:penalty}
	\begin{algorithmic}[1] 
		\Require initial $\mu > 0$, increase factor $\rho>1$
		\Ensure final $\mu$ for penalty function 
		\Repeat
		\State $\ss \gets \arg\min_{\scaleto{\ss}{4.6pt}\,\in {\cal D}} \big(f(\ss)+\dfrac{\mu}{2}\,||c(\ss)||^2\,\big)$
		\Comment{Solve on quantum annealer}
		\If {$c(\ss)>0$}
		\State $\mu \gets \rho\mu$ 
		\EndIf
		\Until{$c(\ss)=0$ or iterations limit reached} \Comment{Other stopping criteria possible}
	\end{algorithmic}
\end{algorithm}

\subsubsection{Augmented Lagrangian method} The augmented Lagrangian method combines the idea of the penalty method (PM) with the method of the Lagrangian multipliers. The Lagrangian multipliers method tries to solve problem \eqref{eq:problem} by constructing the function ${\cal L}(\ss,\ll)=f(\ss)+\ll^T c(\ss)$, where $\ll$ is a vector of coefficients called \textit{Lagrangian multipliers},  and looks at its stationary points, i.e., points where all partial derivatives are zero, including the one with respect to the Lagrangian multipliers $\ll$. These points contain all optimal solutions of the original problem \eqref{eq:problem}. But the Lagrangian multipliers method cannot be directly used with D-Wave, since ${\cal L}(\ss,\ll)$ is not an Ising problem as $\ll$ consists of real numbers, which have to be computed by an iterative method. However, a simple procedure like that of Algorithm~\ref{alg:penalty}, which increases the value of the coefficient at  each iteration, will not work since (i) the exact value of $\ll$ has to be used, unlike PM where the value of $\mu$ just needs to be large enough; (ii) unlike $\mu$, which is a single number, $\ll$ is a vector, which makes its exact estimation more difficult. For that reason, the \textit{Augmented Lagrangian method  (ALM)}, which augments ${\cal L}(\ss,\ll)$ with a quadratic penalty term to produce the unconstrained problem
\begin{equation}
	\mbox{minimize }f(\ss)+\ll^T\, c(\ss)+\dfrac{\mu}{2}\,||c(\ss)||^2,\; \ss\in {\cal D},\label{eq:ALproblem}
\end{equation}
where  $\ll$ is a vector of dimension the number of logical qubits, is a better alternative. It also results in lower values of $\mu$, compared with PM. A pseudocode of a standard implementation of ALM is given in Algorithm~\ref{alg:AL}.

\begin{algorithm}
	\caption{Augmented Lagrangian method}\label{alg:AL}
	\begin{algorithmic}[1] 
		\Require initial $\mu > 0$, initial $\ll$, increase factor $\rho>1$
		\Ensure final $\mu$,  $\ll$ for the augmented Lagrangian function 
		\Repeat
		\State $\ss \gets \arg\min_{\scaleto{\ss}{4.6pt}\in \mathbb{B}^N}\,\big(f(\ss)+\ll^T\, c(\ss) + \dfrac{\mu}{2}\,||c(\ss)||^2\,\big)$
		\Comment{Solve on quantum annealer}
		\If {$c(\ss)\neq 0$}
			\State $\ll \gets \ll + \mu c(\ss)$
			\State $\mu \gets \rho\mu$ 
		\EndIf
		\Until{$c(\ss)=0$ or iteration limit reached} \Comment{Other stopping criteria possible}
	\end{algorithmic}
\end{algorithm}

In the next subsection, we describe the application of the methods to the problem of implementing logical qubits.

\subsection{Applying the methods to QA and logical qubits implementation}
Assume that we are given to solve, on a quantum annealer, the constrained problem determined by equations \eqref{eq:Ising2}--\eqref{eq:constraint}, i.e., the problem
\begin{align}
	\mbox{minimize }\mathit{Is}'(\bm{s'}) &= \sum_{i<j}\frac{J_{ij}}{|C_{ij}|}\sum_{(k,l)\in C_{ij}}\!\!\!s_i^ks_j^l+\sum_{i}\frac{h_i}{|C_i|}\sum_{s_i^k\in C_i}s_i^k \label{eq:ALobjective}\\
	\mbox{subject to }s_x &= s_y \mbox{ for each $(s_x,s_y)\in \bigcup_iE_i$.}\label{eq:ALconstraint}
\end{align}

Converting \eqref{eq:ALobjective}--\eqref{eq:ALconstraint} into an unconstrained problem of  type \eqref{eq:ALproblem} results into the problem
\begin{equation}
	\mbox{minimize }\mathit{Is}^*(\bm{s'}) = \mathit{Is}'(\bm{s'}) + \!\!\!\sum_{(s_x,s_y)\in {\cal E}}\!\!\lambda_{xy}(s_x - s_y)  
	+\frac{\mu}{2}\!\sum_{(s_x,s_y)\in {\cal E}}\!\!(s_x - s_y)^2,  \label{eq:AL2}
\end{equation}
where ${\cal E}=\bigcup_iE_i$. Applying the square formula and using the fact that $s^2=1$ for $s\in\{-1,1\}$, we simplify $\mathit{Is}^*(\bm{s'})$ as
\begin{equation}
	\mathit{Is}^*(\bm{s'}) =  \mathit{Is}'(\bm{s'}) + \!\!\!\sum_{(s_x,s_y)\in {\cal E}}\!\!(\lambda_{xy}s_x - \lambda_{xy}s_y  
	-\mu\, s_xs_y) +\mu|E|.  \label{eq:AL3}
\end{equation}
Note  $\mu|E|$ is a constant that can be ignored when minimizing with respect \mbox{to $\\s'$}.

Formula \eqref{eq:AL3} leads to the following implementation of Algorithm~\ref{alg:AL} for the case of logical qubits given on \cref{alg:AL2}. 
The specific stopping criterion we use is that there are no broken chains or the number of iteration exceeds $20$.

\begin{algorithm}
	\caption{Augmented Lagrangian method}\label{alg:AL2}
\begin{description}
	\item{\algorithmicrequire} Ising model $\mathit{Is}(\bm{s})$, initial $\mu_0 > 0$, initial $\ll_0$, increase factor $\rho>1$
	\item{\algorithmicensure} Final $\mu$,  $\ll$ for the augmented Lagrangian function 
\end{description}
\begin{enumerate}
	\item Find an embedding of $\mathit{Is}(\bm{s})$ and construct the transformed Ising $\mathit{Is}'(\bm{s'})$ using the D-Wave's Ocean software function \texttt{embed\_ising} but setting the parameter \texttt{chain\_strength} to $0$ (so that we can define chain strength by ALM). Let $J'(s_x,s_y)$ and $h'(s_x)$ denote the coefficients in front of $s_xs_y$ and $s_x$ in $\mathit{Is}'(\bm{s'})$, respectively, and let $J^*(s_x,s_y)$ and $h^*(s_x)$ denote the corresponding coefficients of $\mathit{Is}^*(\bm{s'})$.
	\item Set suitable initial values to the parameters of Algorithm~\ref{alg:AL}. In our implementation, we set $\mu\gets \mu_0=1.5$, $\ll\gets\ll_0= \bm{0}$, $\rho\gets 1.1$.
	\item Construct the set ${\cal E}$ of pairs of same-chain qubits that are connected by a coupler. 
	\item Initialize $\mathit{Is}^*(\bm{s'})\gets \mathit{Is}'(\bm{s'})$ and repeat steps (i)-(v) until a stopping criterion has been met.
	\begin{enumerate}
		\item For each $(s_x,s_y)\in {\cal E}$ make the updates
			\begin{algorithmic}
				\State $h^*(s_x) \gets h'(s_x) + \lambda_{xy}$
				\State $h^*(s_y) \gets h'(s_y) - \lambda_{xy}$
				\State $J^*(s_x,s_y) = \mu$
			\end{algorithmic}
		\item Submit $\mathit{Is}^*(\bm{s'})$ to the quantum annealer and get a list of samples (proposed solutions).
		\item Find the sample $\sigma$ resulting in a best solution. 
		\item Update $\ll$ as follows
			\begin{algorithmic}
				\For  {$(s_x,s_y)\in {\cal E}$}
					\State $(\sigma_x,\sigma_y) \gets$ values of $(s_x,s_y)$ from $\sigma$
					\State $\lambda_{xy} \gets  \lambda_{xy} + \mu(\sigma_x-\sigma_y)$
				\EndFor
			\end{algorithmic}
		\item $\mu \gets \alpha\mu$
	\end{enumerate}
\end{enumerate}
\end{algorithm}

ALM is considered a very good choice in practice for a large set of applications \cite{birgin2014practical}, but its application to logical qubits has distinctive aspects that may affect its effectiveness. Such features include the following. 
\begin{enumerate}
	\item The original method has been developed for continuous variables. In that case and under mild conditions, ALM  converges to a global optimum regardless of the initial values of the Lagrange multipliers \cite{doi:10.1137/10081085X}. However, in the case with discrete variables like ours, there are no theoretical results available that guarantee optimality. 
	\item The number of constraints in our case can be quite large (it is equal to the number of variables). ALM has been shown to work well for number of constraints of order hundreds \cite{doi:10.1080/10556780500139690} in the \textit{continuous} case, which gives a hope that it may work in our case as well. But it needs to be checked.
	\item Quantum annealers are stochastic samplers, which means that the outcome of measuring the quantum state at the end of the annealing is random. As a consequence, the result of each iteration depends not only on the parameters $\ll$ and $\mu$, but also on chance. This is further exacerbated by the fact that the annealer is susceptible to noises from the environment, hardware leaks, and other interactions that may substantially affect the outcome of the quantum annealing process \cite{pelofske2022noise}. 
\end{enumerate}

For that reason, it is important to experimentally analyze the practical effectiveness of our algorithm, which we address in the \cref{sec:results}.

\subsection{Using ALM on a class of problems}
So far, we have applied ALM on a single problem to find a representation of the logical qubits for that particular problem. Here we propose a modification that allows ALM to construct implementations of logical qubits that can then be used to solve a set of problems from a given class. In order to allow that, the  problems in the class should have the same number of variables (logical qubits). The algorithm is similar to \cref{alg:AL2}, but we solve each problem in the set separately and, when we update the Lagrangian coefficients $\ll$, we take into account the violated constrains (broken chains) in \textit{all} of the solutions. The details are given in \cref{alg:ALM-opt}.

\begin{algorithm}
	\caption{ALM for a set of problem of the same size}\label{alg:ALM-opt}
	\begin{description}
		\item{\algorithmicrequire} A set ${\CMcal S}$ of Ising models $\mathit{Is}_1(\bm{s}), \dots, \mathit{Is}_k(\bm{s})$, initial $\mu_0 > 0$ and $\ll_0$, increase factor $\rho>1$
		\item{\algorithmicensure} Final $\mu$,  $\ll$ for the set of Ising models 
	\end{description}
	\begin{description}
		\item[1.] Find an embedding of a complete graph of $\mathrm{dim}(\bm{s})$ vertices and use it to embed
		each graph of ${\CMcal S}$, and construct the set ${\CMcal S}'$ of transformed Ising models $\mathit{Is}'_1(\bm{s}), \dots, \mathit{Is}'_k(\bm{s})$		 
		using Ocean's \texttt{embed\_ising} function while setting the parameter \texttt{chain\_strength} to $0$. Let $J'$, $h'$, $J^*$, and $h^*$ be defined as in \cref{alg:AL2}.		 
		\item[2--3.] Set $\mu$, $\ll$, $\rho$, and ${\cal E}$ as in \cref{alg:AL2}. 
		\item[4.] Initialize $\mathit{Is}_i^*(\bm{s'})\gets\mathit{Is}_i'(\bm{s'})$ for $1\leq i\leq k$ and denote ${\CMcal S}^*=\{\mathit{Is}^*_1(\bm{s}), \dots, \mathit{Is}^*_k(\bm{s})\}$. Repeat steps (i)-(v) until a stopping criterion has been met.
		\begin{description}
			\item[(i)] Update $\bm{h^*}$ and $\bm{J^*}$ as in \cref{alg:AL2}.
			\item[(ii)] Submit each problem from ${\CMcal S}^*$ to the quantum annealer and get a list of samples for each of them.
			\item[(iii)] Find the sample $\sigma_i$ resulting in the best solution of $\mathit{Is}_i^*$ for $1\leq i\leq k$. 
			\item[(iv)] Update $\ll$ as follows
			\begin{algorithmic}
				\For  {$(s_x,s_y)\in {\cal E}$}
				\State $(\sigma_x^i,\sigma_y^i) \gets$ values of $(s_x,s_y)$ from $\sigma_i$ for $1\leq i\leq k$
				\State $\lambda_{xy} \gets  \lambda_{xy} + \mu\sum_{i=1}^{k}(\sigma^i_x-\sigma^i_y)/k$
				\EndFor
			\end{algorithmic}
			\item[(v)] $\mu \gets \alpha\mu$
		\end{description}
	\end{description}
\end{algorithm}

\section{Results}\label{sec:results}

\subsection{Experimental set-up}
We test the proposed method on instances of the maximum clique problem. A \textit{clique} is a graph that is fully connected, i.e., such that there is an edge between each pair of vertices. A \textit{maximum clique (MC)} of a graph $G$ is a subgraph of $G$ that is a clique of maximum size. The \textit{maximum clique problem (MCP)} asks to find a maximum clique in $G$ and it is an $NP$-hard problem \cite{Karp72_NP}.

For constructing input graphs for  MCP, we generate Erd\H{o}s–R\'{e}nyi random graphs using probability parameter $p=0.5$ and number of vertices $n\in\{50,75,100\}$. For each combination $(n,p)$ of values, we generate five random $G(n,p)$ graphs.

In order to solve an MCP on a D-Wave annealer, we have to encode it as a QUBO or Ising problem. For this end, we define a binary variable $x_i\in\{0,1\}$ for each vertex $i$ of $G$ that indicates whether $i$ is included in the proposed clique or not. Then, the maximum clique problem can be formulated as
\begin{equation}
	    \mbox{minimize } Q = -A\,\sum_{i \in V} x_i + B\!\sum_{(i,j) \in \overline{E}} x_i x_j,
	\label{eq:MC}
\end{equation}
where $\overline{E}$ is the set of edges of the complement of $G$ and the positive constants $A$ and $B$ should satisfy $A<B$  \cite{Lucas2014}. In \eqref{eq:MC}, the first term encodes the objective function, which is the clique size (weighted by $-A$), and the second one encodes the constraint function as a penalty, which takes value at least $B=B\cdot 1$ if the set of vertices $i$ with $x_i=1$ does not form a clique. In our implementation, we choose $A=1$ and $B=2$. Since problem \eqref{eq:MC} is a QUBO while Algorithm~\ref{alg:AL2} requires an Ising representation, we transform \eqref{eq:MC} into an Ising problem using D-Wave's Ocean function \texttt{qubo\_to\_ising}.

For our experiments, we use \texttt{Advantage\_system4.1} quantum annealer available through the D-Wave's Leap quantum cloud service. The annealing settings we use are 1000 for the number of samples and 100\,$\mu$s for the annealing time. The chain strength for the ALM and PM implementations are determined as a result of running Algorithms~\ref{alg:AL2} and \ref{alg:penalty}, respectively, and for the default logical qubits implementation we use as a chain strength the value computed by D-Wave Ocean's recommended \texttt{uniform\_torque\_compensation} function,
which determines a chain strength that usually results in a small number of broken chains, with a  prefactor set to the default value of $1.414$. For handling broken chains, we use the default \texttt{majority\_vote} method, which assigns a value ($+1$  or $-1$) to a logical qubit equal to the the most common value of its physical qubits. 

Next, we compare the performance of ALM against PM and the default D-Wave method with uniform torque compensation.

\subsection{Number of chain breaks per iteration}
\begin{figure}
	\centering
	\includegraphics[width=0.49\textwidth]{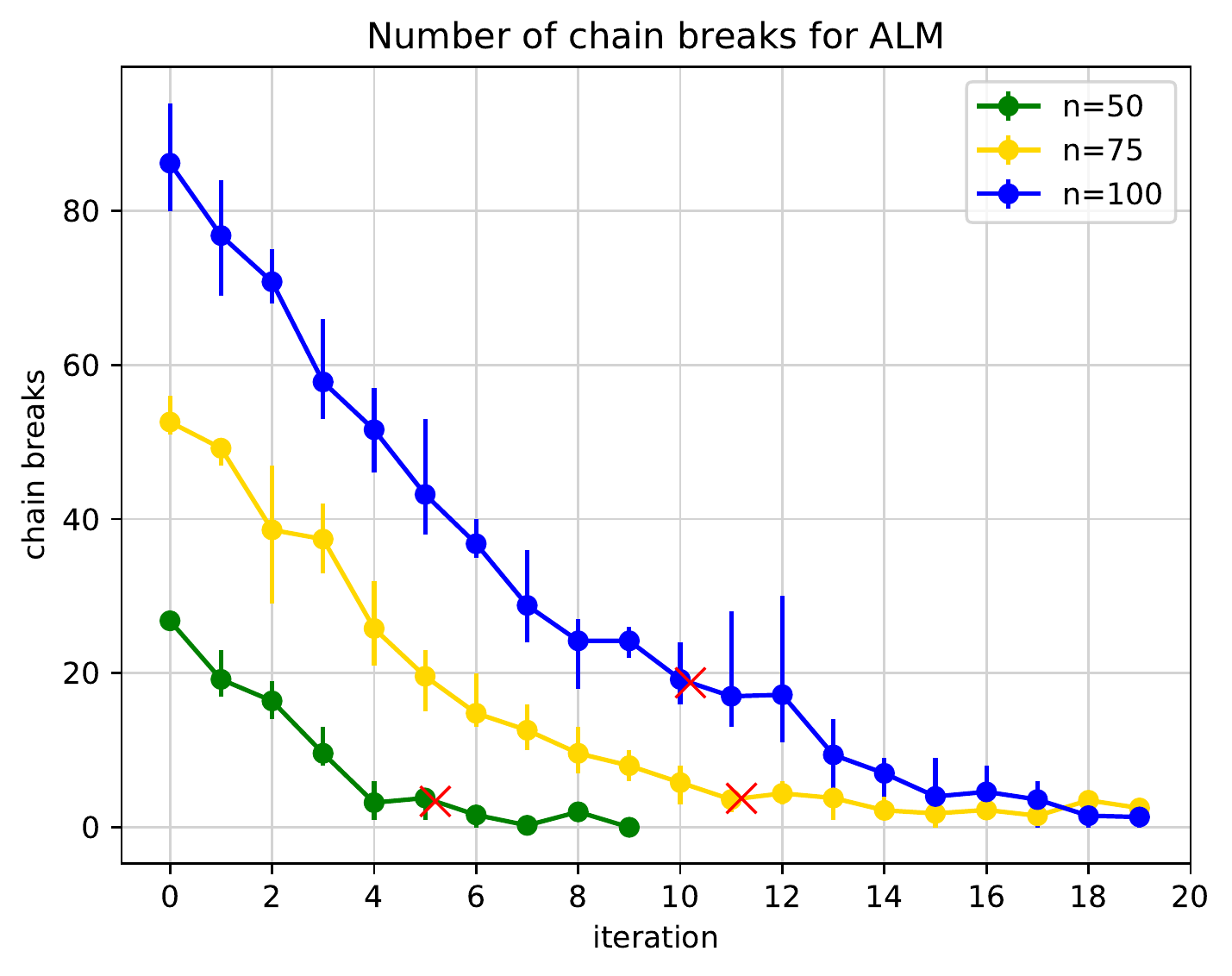}\hfill
	\includegraphics[width=0.49\textwidth]{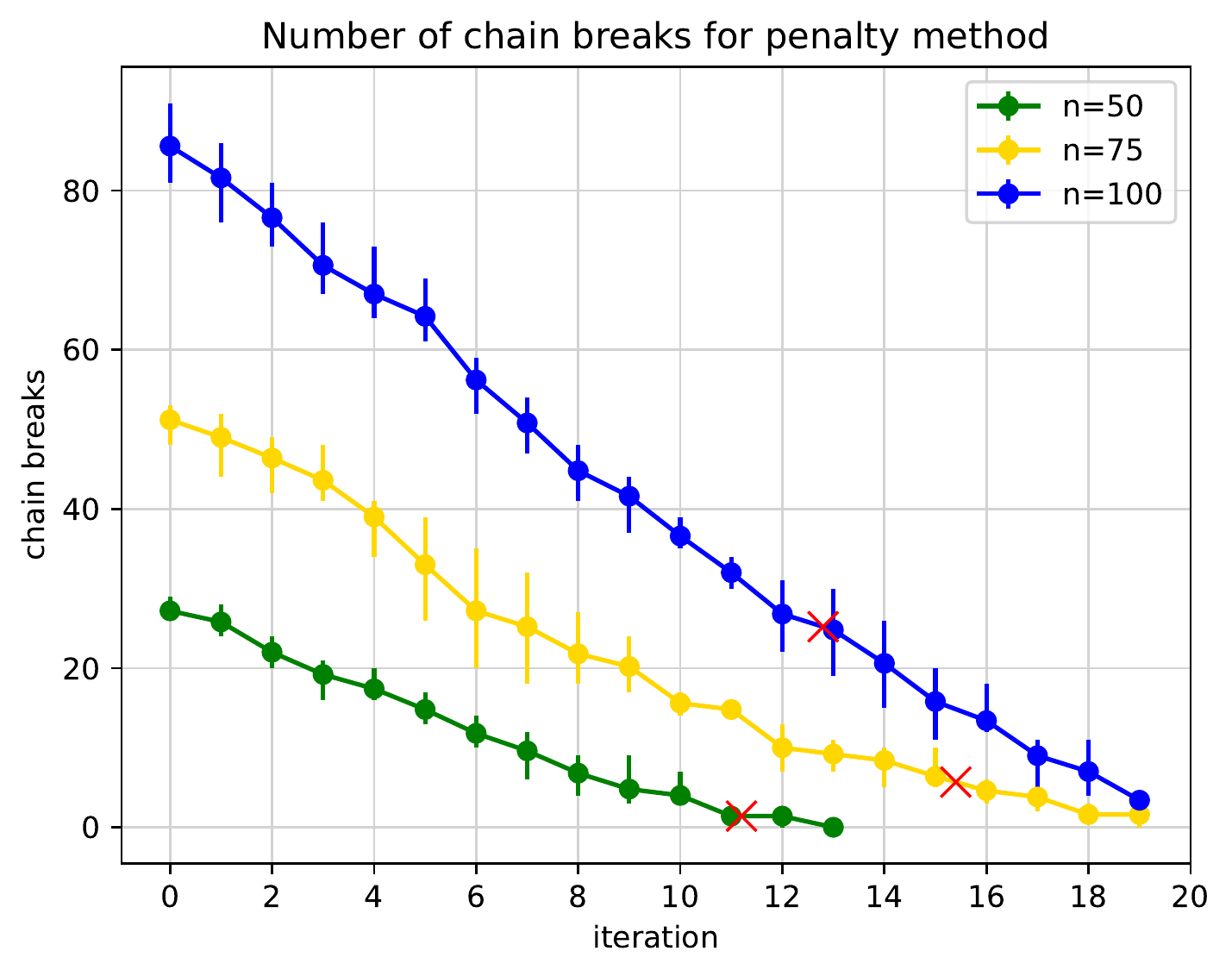}
	\caption{Number of chain breaks per iteration for  ALM  (left) and for PM (right) for random graphs of various sizes. Dots correspond to the average values and the bars indicate the range of values for the corresponding iteration, while ``{\color{red}$\times$}'' indicates the average iteration number at which the highest value for the clique number is obtained.}\label{fig:ch_breaks}
\end{figure}

As the goal of the  algorithms we described is to deal with  chain breaks (CB) in logical qubits, our first goal is the examine how the number of CB changes during the iterations and how ALM compares with the PM implementation. \Cref{fig:ch_breaks}  shows the average number of CB per iteration on the left, and that number for PM on the right. Red "X" indicates the optimal iteration in terms of the average value of the MC number. We observe that the number of CB is quite large in the beginning, and that number goes down to zero in all cases, more steeply in the beginning and flattening out for the ALM method, while the reduction is more gradual and with roughly constant slope for PM. Also, for both methods and for all graph sizes, the best value for the maximum clique is found a few iterations before the number of CB gets down to zero, rather than at the last iteration. That can be explained with the trade-off between reducing the number of CB and increasing the value of the penalty parameter $\mu$. Specifically, on one side, the elimination of the broken chains makes the Ising model a better match to the original problem \cref{eq:Ising1}. On the other side, each iteration increases the value of $\mu$ by a factor of 1.1, and such an increase negatively impacts the accuracy of the Ising model \cref{eq:AL3} solution. Hence, the best solution may be found at an iteration at which the number of CB is not zero, but $\mu$ is only moderately large.

Finally, comparing the number of iterations between ALM and PM to get to a best solution, we see ALM uses fewer iterations than PM in all three cases (5.2 vs 11.2 for $n=50$, 11.2 vs 15.4 for $n=75$, and 10.2 vs 12.8 for $n=100$). Note that ``best'' here means the best solution over all iterations of a particular run. While, for $n=100$ case, 10.2 vs 12.8 looks like ALM has only a modest advantage over PM in terms of number of iterations,  the difference in the quality of solution is much more substantial, as we will see in \cref{sec:MC}.

\subsection{Values of the penalty factor}
\begin{figure}
	\centering
	\includegraphics[width=0.4\textwidth]{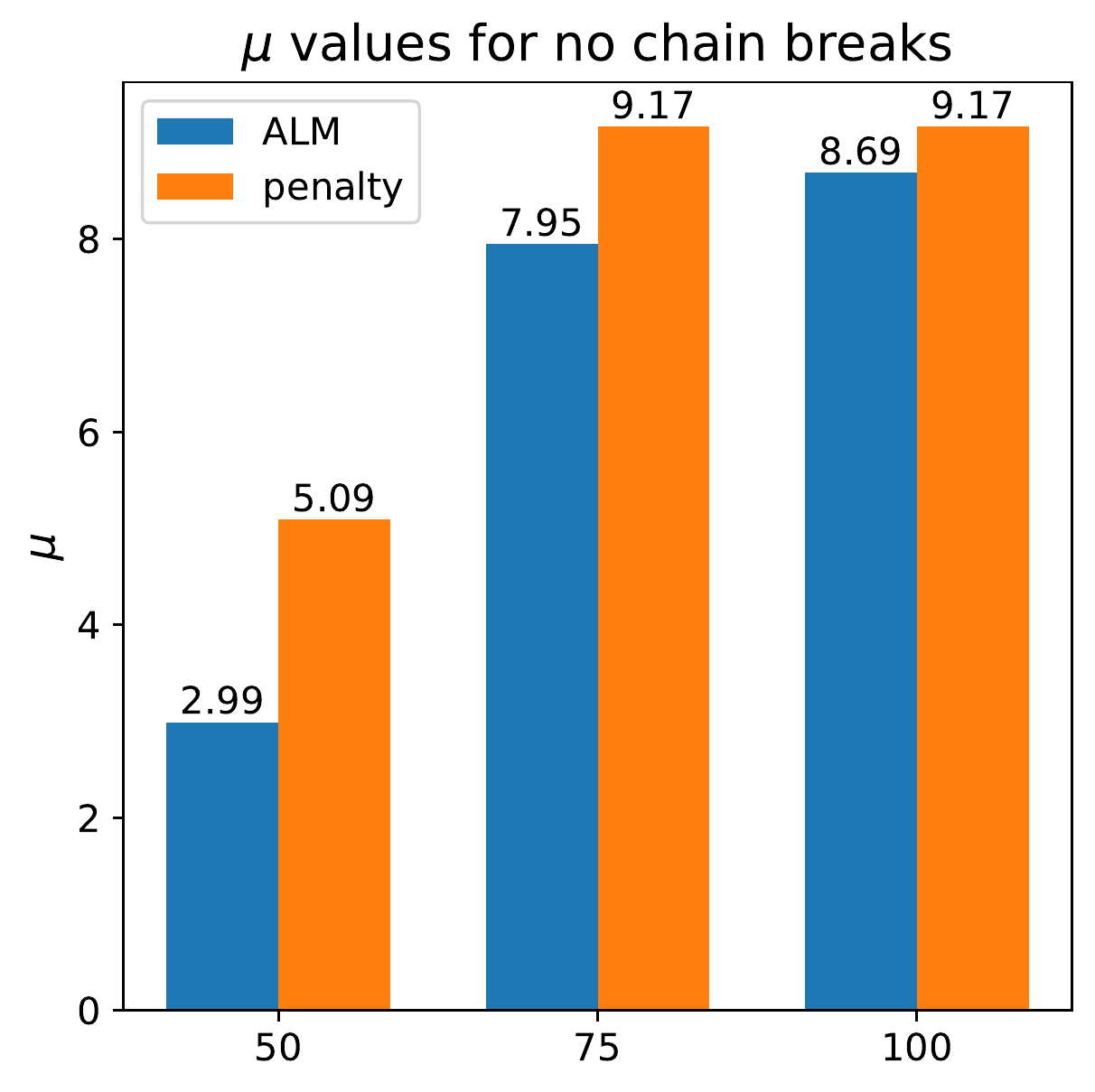}\hspace{1cm}
	\includegraphics[width=0.4\textwidth]{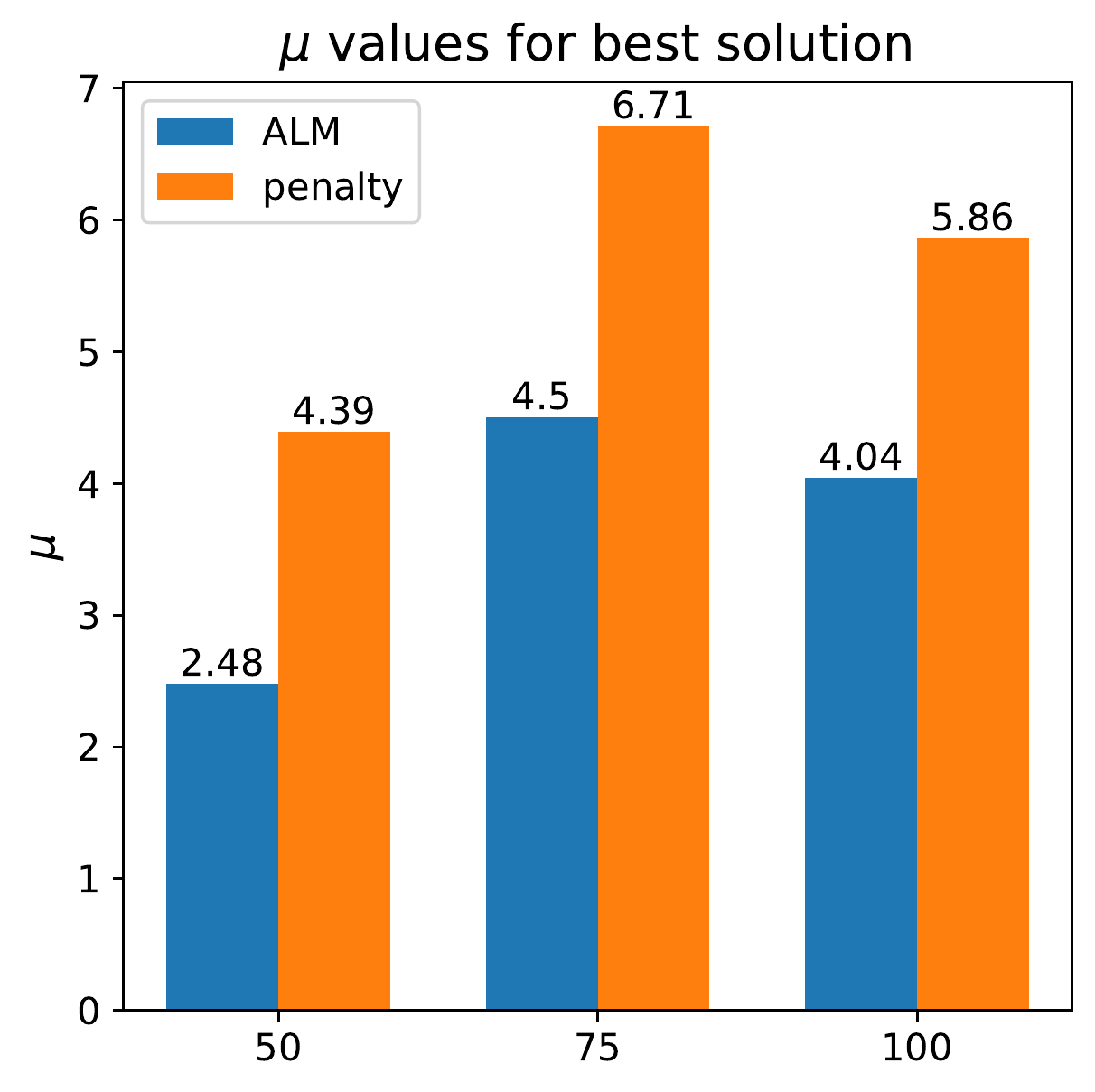}
	\caption{Average values of the penalty factor, $\mu$, at the completion of ALM and PM, i.e., when the number of chain breaks becomes zero (left), and the corresponding values at the iterations where a best value for the MC number is found (right).}\label{fig:mu_values}
\end{figure}

Next we compare the values of the penalty factor, $\mu$, that ALM and PM obtain at the completion of the algorithms, as well as those values at the iteration where the best value of the MC number is achieved. 
The results are shown in \cref{fig:mu_values}.
We can see that ALM gets consistently better (lower) values, with reductions over PM $41\%$, $13\%$, $5\%$, for $n=50$, $75$, and $100$ and the last-iteration case, and reductions of $44\%$, $33\%$, $31\%$, for the best-MC iteration.

\subsection{Values of the largest clique numbers found}\label{sec:MC}
Probably the most important comparison criterion is whether and how much the new algorithm can help to get better solutions than the previous ones. In this subsection, we show comparison between ALM, PM, and the standard default method (SM) offered by the D-Wave software. The default \texttt{uniform\_torque\_compensation} (UTC) method by D-Wave's Ocean software, used by SM, is a heuristic procedure that assigns a chain strength (penalty factor) by computing the root mean square (RMS) of the quadratic couplers of the Ising or QUBO model, multiplied by the square root of the average degree of the Ising graph vertices and a constant prefactor with default value of $1.414$.

\begin{figure}
	\centering
	\includegraphics[width=0.5\textwidth]{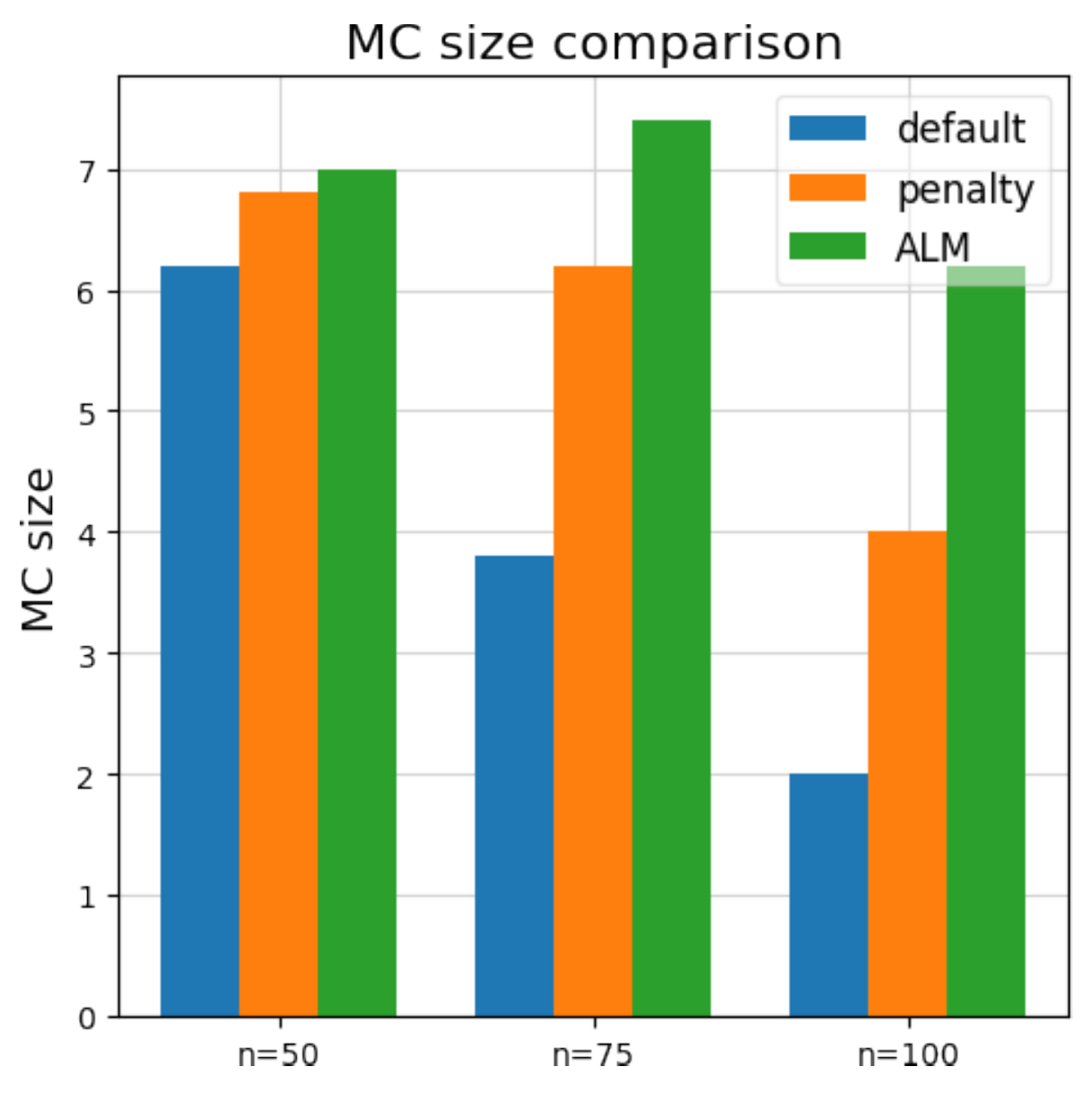}\qquad
	\captionof{figure}{Comparison between default, penalty, and ALM methods on random test graphs of various sizes.}
	\label{fig:MC_sizes}
\end{figure}

\cref{fig:MC_sizes} shows the results of comparing SM, PM, and ALM on random test graphs of various sizes. Each bar shows the average MC size found by the corresponding algorithm and graph size. For instances where no valid clique is returned, we assign MC size of zero. We can see that ALM consistently gets better values for the MC number than the other methods. The biggest difference is for the largest size ($n=100$), where the average clique size found by ALM is twice larger than the one found by PM and more than three times larger than the one found by SM. The large difference in performance between ALM and PM is  due to the fact that ALM arrives at smaller values of $\mu$, which helps in the accuracy of the quantum annealing, and, additionally, it uses Lagrangian coefficients $\lambda$ to help enforce the chain constraints, while PM relies only on $\mu$. 

Comparing SM and PM we observe that the default values for $\mu$ used by SM are relatively good, but only for the smallest and easiest case $n=50$. The performance of SM relative to PM deteriorates as the problem gets more difficult with the size increase. That means that SM and the UTC heuristic  don't use the full potential of PM and improvement may be possible with a more sophisticated and better heuristic or by just using an iterative algorithm of the type of \cref{alg:penalty} for penalty factor estimation.

\subsection{ALM on a class of problems}
In order to use \cref{alg:ALM-opt} on a set of problems, all of the problems in the set need to have the same number of variables, which in the case of the MC problem means that the input graphs have to be of the same size. In our experiments, we choose our set ${\CMcal S}_n$ to be the set of
Erd\H{o}s–R\'{e}nyi random graphs $R(n,p)$ with probability parameter $p=0.5$ and a fixed $n$. As in the other experiments in this section, $n$ takes values in $\{50,75,100\}$.  
To compute values for $\mu$ and $\ll$, for each $n\in\{50,75,100\}$, we apply  \cref{alg:ALM-opt} on a set of ten randomly chosen graphs from ${\CMcal S}_n$ and store the values for later use.

\begin{figure}
	\centering
	\includegraphics[width=\textwidth]{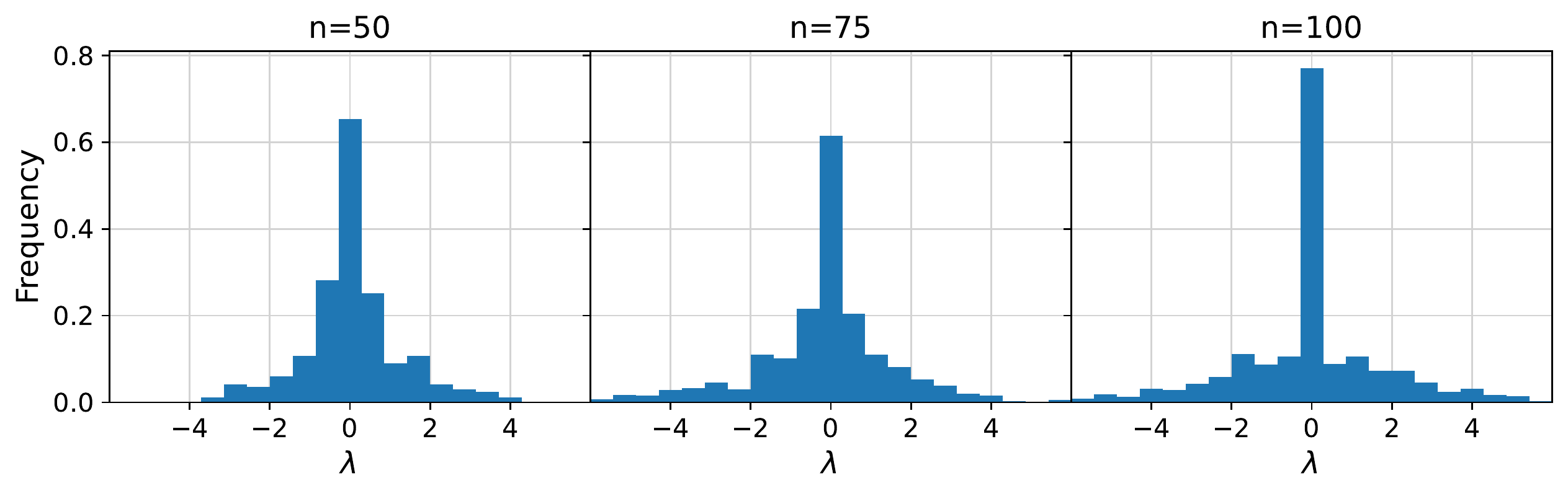}
	\caption{Distribution of the $\lambda$ values for the \almset algorithm. The \texttt{density} parameter of the \texttt{matplotlib.pyplot.hist} method was used to normalizes bin heights so that the integral of the histogram equals one. }\label{fig:lam_distribution}
\end{figure}

To give an idea about a typical range of values for the Lagrangian coefficients $\ll$, we round their values to the closest integer and put them in bins, for each of the three graph sizes. The resulting histograms are shown in \cref{fig:lam_distribution}. We can see that zero is the most common value and that most of the values are between $[-4,4]$. 

\begin{figure}
	\centering
	\begin{minipage}{.47\textwidth}
		\centering
		\includegraphics[width=\textwidth]{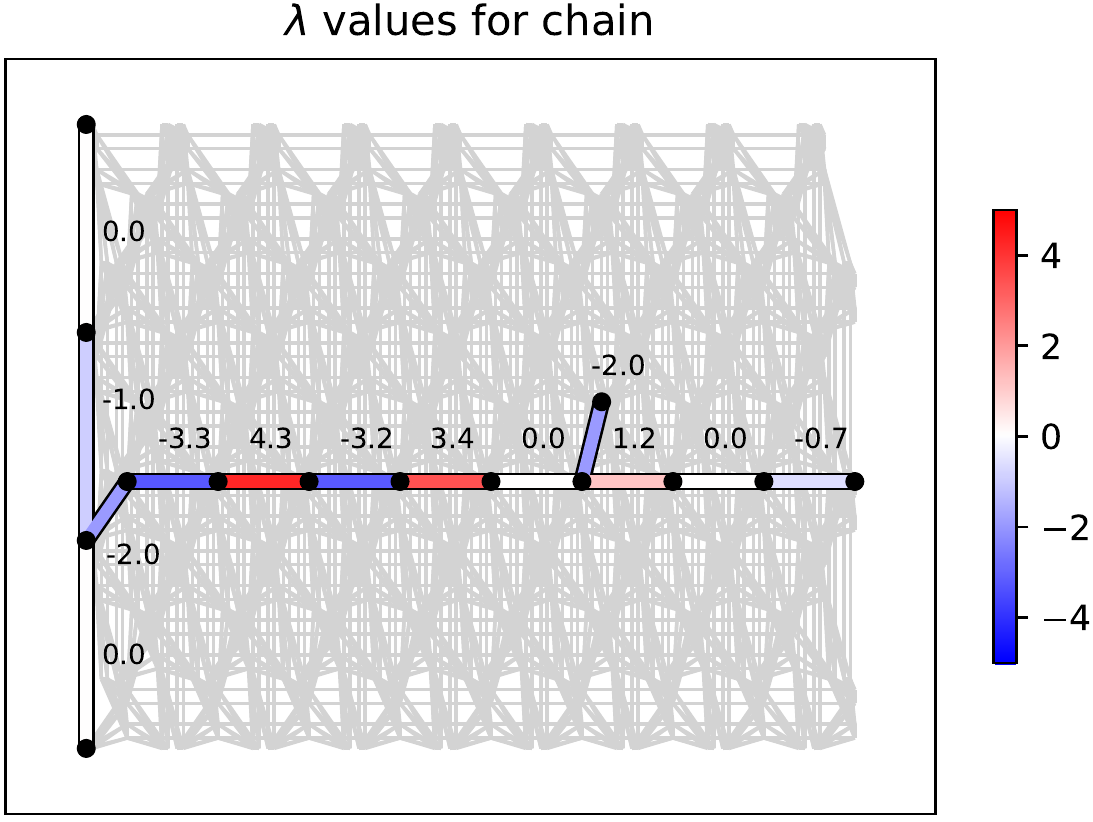}
		\caption{Values of the $\lambda$ parameter computed for a specific randomly picked chain from the embedding for $n=100$.}\label{fig:lam_chain}
	\end{minipage}%
	\hfill
	\begin{minipage}{.47\textwidth}
		\centering
		\includegraphics[width=\textwidth]{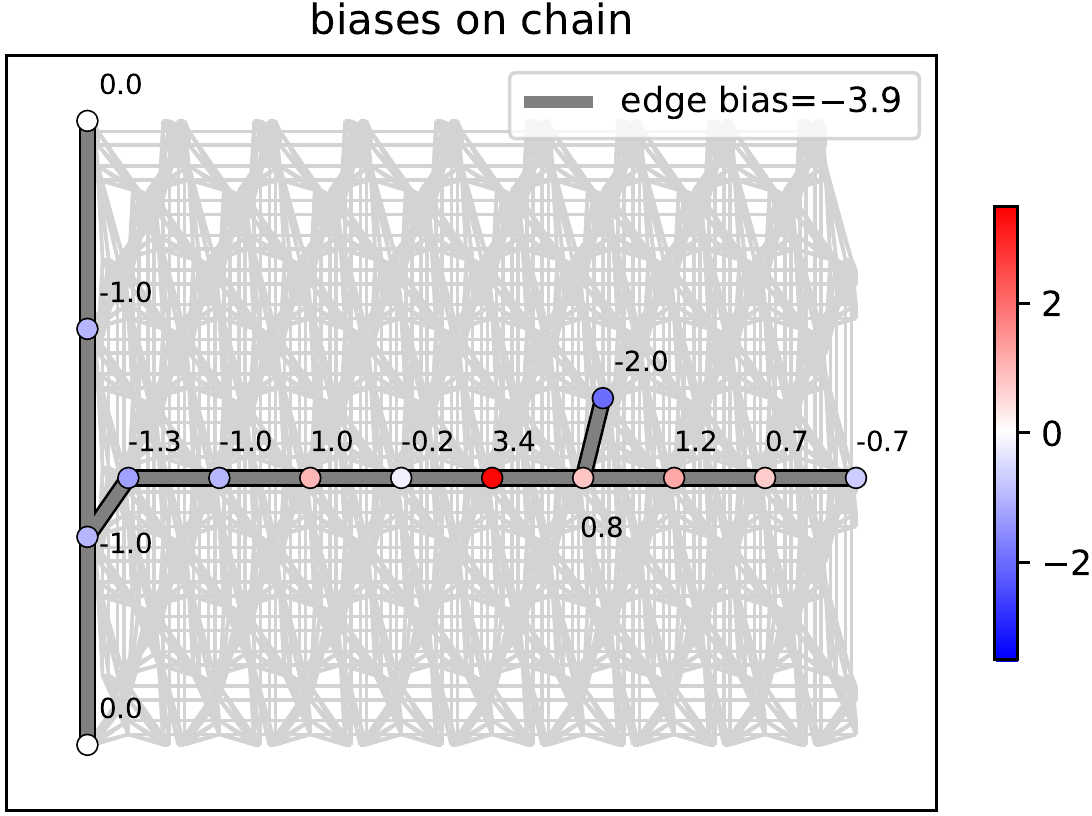}
		\caption{Linear and quadratic biases for the same chain as in \cref{fig:lam_chain}.\\}\label{fig:biases}
		
	\end{minipage}
\end{figure}

Furthermore, to illustrate the distribution of  $\lambda$ values on the edges of a typical chain, we picked one chain from the embedding for $n=100$ and plotted on each edge of that chain the corresponding $\lambda$ value, as shown on \cref{fig:lam_chain}. While there are both positive and negative values on the edges, the signs are irrelevant and only the magnitudes matter, because the signs depend on the directions of the corresponding edges, which are arbitrarily oriented. These $\lambda$ values, along with the $\mu$ value, which in this particular case is $-3.9$, are used to compute the additional linear and quadratic biases on qubits and couplers corresponding to the chain constraints, using equation \cref{eq:AL3}. The resulting additional biases are illustrated on \cref{fig:biases}. As one might expect, there are portions of the chain with mostly positive biases and portions where the negative biases dominate. Note that these linear biases are not necessarily the ones in the final Ising problem sent to the quantum annealer as in addition to the chain biases there are also linear biases coming from the problem Ising \cref{eq:ALobjective}, and the two biases are added together. On the other hand, the quadratic chain biases, set to $-3.9$ for that particular chain, are the final ones.

In order to compare the accuracy of the algorithms, we run them on different sets ${\CMcal S}_n$ of random graphs and compare the sizes of the MCs found. The first of these algorithms is the default D-Wave algorithm, SM, considered in the previous subsection. The second algorithm, \almset,  uses the representations of the logical qubits defined by the stored values of $\mu$ and $\ll$. The difference between ALM from the previous subsection and \almset is that the former uses a number of iteration to arrive at suitable values for $\mu$ and $\ll$ for the particular graph, while the latter takes just a single annealing call. The third algorithm we compare, \almsetpl, adds a single augmented Lagrangian iteration after running \almset. The rationale for this is that the stored values of $\mu$ and $\ll$ are optimized for the class ${\CMcal S}_n$ as whole, but they may not be well-suited for any particular graph in that class. We conjecture that, by using $\mu$ and $\ll$ as initial values, a single iteration may help to adjust the logical qubit representation so that they better match the particular features of the specific graph, and result in better results than \almset. 

\begin{figure}
		\centering
		\includegraphics[width=0.5\textwidth]{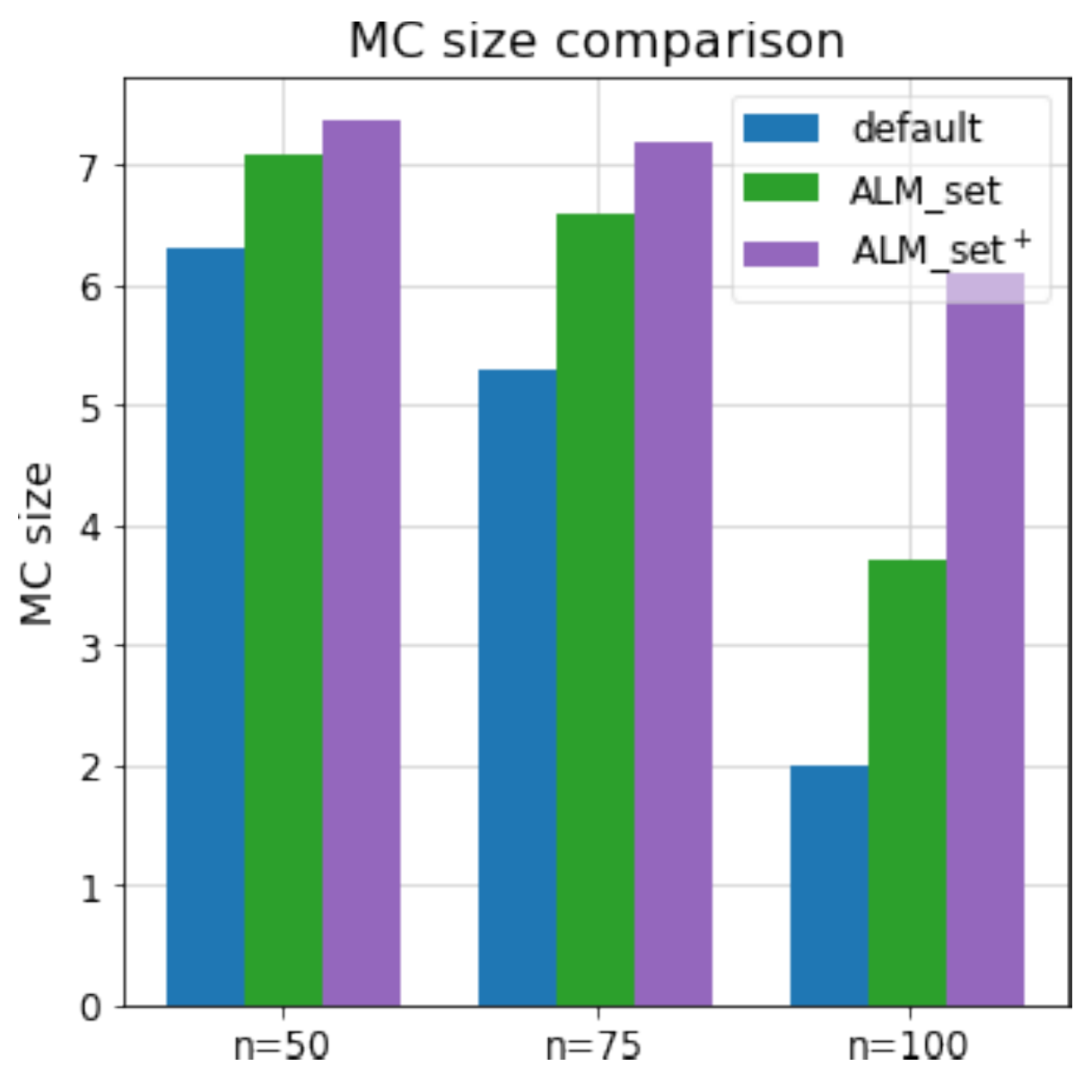}
		\captionof{figure}{Comparison between default, \almset and \almsetpl methods on sets of random graphs of given sizes.}
		\label{fig:opt_vs_default}
\end{figure}

\cref{fig:opt_vs_default} shows the results of this experiment. 
Since \almsetpl takes the better of two MC values, one from the \almset run and one from the single augmented Lagrangian iteration, in order to make the comparison fair, we also run SM and \almset twice and take the better of the two MC values.
We observe that storing $\mu$ and $\ll$ values (\almset) does lead to significant improvements over SM. However, despite the improvement, \almset results for $n=100$ are significantly worse than ALM results, \cref{fig:MC_sizes}. Adding the extra iteration (\almsetpl), however, leads to a significant improvement and MC values that roughly match the ones from ALM. 

\vspace{1cm}


\section{Conclusion}
In this paper, we showed that the augmented Lagrangian method can be used to determine a representation of logical qubits for a given Ising problem in terms of quadratic and linear biases on the chain couplers and qubits by an iterative optimization procedure. We experimentally analyzed the method on the maximum clique problem on random graphs of various sizes and observed that our proposed method outperforms quadratic-penalty based methods. With increasing the graph size and the difficulty, the performance gap becomes more significant. We also propose a modification that allows values for the chains' biases for entire classes of graphs to be generated and saved. While this method outperforms the method that uses default chain weights, the version of the method that is adding a single augmented Lagrangian iteration matches the performance of the full augmented Lagrangian version (with 5--11 iterations).

The paper leaves several open problems for future research. Our formulation assumed that, for optimization problem that originally have had constraints, they have already been included in the objective function as penalty terms. But since we are using ALM to find representations of the logical qubits, we can use the same iterations to find better representations of the constraints as well. The problem we used in our experiments, the maximum clique problem, has a relatively low penalty factor of two for its constraint, but many other optimization problems need much larger penalty factors and using the ALM method can make a difference. Another possible improvement is use an individual penalty factor $\mu_i$ for the $i$-th logical qubit, than a single one as in our implementation.


\section*{Acknowledgments}
\label{sec:acknowledgments}
This work was supported by grant number KP-06-DB-11 of the Bulgarian National Science Fund and by 
the U.S. Department of Energy through the Los Alamos National Laboratory. Los Alamos National Laboratory is operated by Triad National Security, LLC, for the National Nuclear Security Administration of U.S. Department of Energy (Contract No.\ 89233218CNA000001).

\section*{References}
\bibliographystyle{plain}

\end{document}